\documentclass[epj]{svjour}
\usepackage{graphicx}
\usepackage{array}
\usepackage{cite}
\usepackage{amssymb}
\usepackage{hyperref}
\begin{document} \title{Non-linear and
    quantum optics of a type II OPO containing a birefringent element
    Part 2~: bright entangled beams generation}

\titlerunning{Non-linear and quantum $\dots$~Part 2~:
  bright entangled beams generation}
\author{L. Longchambon\and J. Laurat\and T. Coudreau \and C. Fabre  }%

\authorrunning{L. Longchambon \emph{et. al.}}

\institute{Laboratoire Kastler Brossel, Case 74, UPMC, 4, Place
  Jussieu, 75252 Paris cedex 05, France }
\mail{coudreau@spectro.jussieu.fr}

%\maketitle
\date{\today}%{Received: date / Revised version: \textbf{\today}}
% The correct dates will be entered by Springer
%
\abstract{We describe theoretically the quantum properties of a
type-II Optical Parametric Oscillator containing a birefringent
plate which induces a linear coupling between the orthogonally
polarized signal and idler beams and results in phase locking
between these two beams. As in a standard OPO, the signal and
idler waves show large quantum correlations which can be measured
experimentally due to the phase locking between the two beams. We
study the influence of the waveplate on the various criteria
characterizing quantum correlations. We show in particular that
the quantum correlations can be maximized by using optimized
quadratures.
 % end of PACS codes
} %end of abstract
\maketitle

\section{Introduction} \label{intro} The domain of quantum information with intense
light beams is becoming more and more studied as practical
applications are developed \cite{teleportkimble,grangier,leuchs}.
A large number of experiments rely on so-called EPR beams, that is
a pair of fields containing quantum correlations on two different
non commuting observables \cite{epr}. Several methods have been
used to date. One way to generate them is to mix two independent
squeezed beams  produced by two independent OPOs below threshold
\cite{teleportkimble,eprbachor}, or to use third order non
linearity in a fiber \cite{leuchs}. Another configuration consists
in using a non degenerate OPO below threshold
\cite{eprkimble,eprtaiyuan,eprpolzik}. We propose here a new
method where two bright entangled beams are produced directly at
the output of a non degenerate OPO pumped above threshold.

Optical Parametric Oscillators are very efficient sources of  non
classical light
\cite{taiyuan,kimblevidesq,mertz,mlynek,zhang,mescond}. In
particular, it has been shown experimentally that the intensities
of the signal and idler beams exhibit very large quantum
correlations \cite{taiyuan,mertz,mescond}. It is also predicted
that the phase of the signal and idler should show quantum
anticorrelations \cite{reynaud}. However, the measurement of these
phase anticorrelations is not possible in a standard OPO in which
frequency degenerate operation occurs only accidentally and is
very sensitive to thermal and mechanical drifts. Thus, in
practical experiments, the frequencies of the signal and idler
beam are different. This implies that the relative phase between
the two fields has a rapidly varying component. Furthermore, even
if the device is actively stabilized on the frequency degeneracy
working point \cite{pfister}, this relative phase between signal
and idler undergoes a diffusion process \cite{courtois}, similarly
to the phase of a laser field.  In order to be able to measure
experimentally these phase anticorrelations, one can use a phase
locked OPO : one possible technique proposed in \cite{wong} is to
use a birefringent plate inserted inside the OPO cavity and making
an angle with the axis of the non-linear crystal which leads to
self-phase locking. This phenomenon has been studied in detail in
\cite{fabre,part1} at the classical level. It is the purpose of
this paper to evaluate the influence of the phase locking
mechanism on the quantum properties of the system.

The paper is organized as follows. In section \ref{sec:base}, we
give the equations describing the quantum fluctuations of the
system. We study in section \ref{sec:variances} the noise spectra
 of the amplitude quadratures difference and phase quadratures sum
showing quantum noise reduction, \emph{i.e} that these variances
are below what would be observed for independent coherent states.
In section \ref{sec:entanglement}, various entanglement criteria
are discussed, namely EPR correlation properties (section
\ref{sec:epr}), inseparability criterion (section
\ref{sec:inseparability}) and teleportation fidelity (section
\ref{sec:fidelity}). Finally, section \ref{sec:opt} is devoted to
the case of non-orthogonal quadratures which enable us to increase
dramatically the violation of the inseparability criterion.

\section{Basic equations}
\label{sec:base}

In order to calculate the fluctuations for the signal and idler
fields when the OPO is pumped above threshold, we apply the
input-output linearization technique described \emph{e.g.} in
\cite{fabre90}. We first determine the classical stationary
solutions of the evolution equations : these solutions are denoted
$A_i^{(s)}$ where $i=0,1,2$ corresponds to the pump, signal and
idler fields. We then linearize the evolution equations around
these stationary values by setting~: $A_i = A_i^{(s)} + \delta
A_i, \,i=0,\, 1,\, 2$.  Introducing the input fluctuations in all
the modes, we obtain a set of linear coupled equations for the
fluctuations of the different fields. By solving these equations
in the Fourier domain, one can calculate the individual noise
spectra as well as the correlations between the different
quadratures.

\begin{figure}[h]
\centerline{\includegraphics[width=0.75\columnwidth]{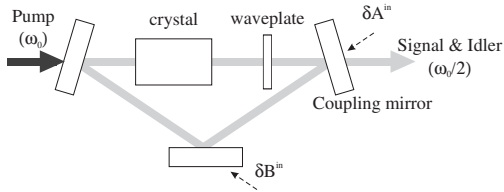}}
\caption{Ring cavity OPO with the fluctuations corresponding to
the fluctuations coming through the input mirror ($\delta
A_i^{in}$) and the fluctuations due to the losses ($\delta
B_i^{in}$).\label{ringcav}}
\end{figure}

Let us recall the system under study which has been discussed in
detail in \cite{part1}. We consider a $\chi^{(2)}$ medium placed
inside a ring optical cavity doubly resonant for signal and idler
(fig.~\ref{ringcav}).
Type-II phase-matching results  in orthogonally polarized signal
and idler fields. The reflection coefficients for signal and idler
are assumed to be equal in module : $r_j = r \exp (i \zeta_j), \,
j=1,2$. We assume also that the cavity has a high finesse for
these modes so that we can write $r = 1 - \kappa$ with $\kappa \ll
1$ ; the transmission is thus $t\approx \sqrt{2\kappa}$. Finally,
the cavity is characterized by the round-trip losses for the
signal and idler waves (crystal absorption, surface scattering),
$\mu$, and we define a generalized reflection coefficient~: $r' =
r (1- \mu) \approx 1 - ( \mu + \kappa ) = 1 - \mu' $.

Let us denote  $l$ the crystal length and $L$ the free propagation
length inside the cavity. The crystal indices of refraction are
$n_1$ and $n_2$ respectively for the signal and idler waves. The
waveplate is rotated by an angle $\rho$ with respect to the
crystal neutral axes. In contrast to \cite{part1}, we assume here
that this angle is small.

The crystal is birefringent and can be characterized by its mean
index $\bar n$, birefringence $\theta$ and effective nonlinear
coupling constant $g$~:
\begin{eqnarray}
\bar n &=& \frac{n_1 + n_2}{2}\\
\theta &=& \frac{\omega_0}{2c} (n_2-n_1) l \\
g &=& l \chi^{(2)} \sqrt{\frac{\hbar \omega_0 \omega_1
\omega_2}{2c^3 \varepsilon_0 n_0 n_1 n_2}}
\end{eqnarray}
where $\omega_0/2$ is the common frequency of signal and idler.

The mean round trip phase-shift is $\delta= \frac{\omega_0}{2c} (n
e + \bar n l +L)+\bar \zeta$ where $\bar \zeta$ is the mean mirror
birefringence, $\bar \zeta = \frac{\zeta_1+\zeta_2}{2}$. We assume
that the signal and idler modes are close to resonance so that the
corresponding round trip phase shifts~
\begin{equation}
\delta_{1,2} = \delta \pm \frac{\theta + \zeta_2-\zeta_1}{2} \mp
\psi
\end{equation}
where $\psi$ is the birefringence of the waveplate, are close to
an integer multiple of $2\pi$~:
\begin{equation}
\delta_{1,2} = 2 \pi p_{1,2} + \Delta_{1,2}
\end{equation}
where  $p_{1,2}$ are integers and $\Delta_{1, 2}$ are small
detunings.

In this case, the classical evolution equations given in
\cite{part1} become ~:
\begin{equation}
\begin{array}{>{\displaystyle}c>{\displaystyle}c>{\displaystyle}l}
\tau \frac{d A_1^{(s)}}{dt} &= & A_1^{(s)} (-\mu' + i \Delta_1) +
g A_0^{(s)} A_2^{(s)\ast} + i 2 \rho
e^{i (\theta-\psi)} A_2^{(s)}\\
\tau \frac{d A_2^{(s)}}{dt} &=& A_2^{(s)}(-\mu' + i \Delta_2) +
gA_0^{(s)} A_1^{(s)\ast} + i 2 \rho e^{-i(\theta-\psi)} A_1^{(s)}\\
 A_0^{(s)}& =&  A_0^{in} - \frac{g}{2} A_1^{(s)} A_2^{(s)}
\end{array}
\end{equation}
where $A_0^{in}$ is the input pump amplitude at frequency
$\omega_0$.

The stationary solution of these equations, corresponding to the
phase locked regime, can be found only within a given range of
detunings $(\Delta_1,\, \Delta_2)$ which has been detailed in
\cite{part1}.

The evolution equations for the fluctuations then are~:
\begin{equation}
\begin{array}{>{\displaystyle}c>{\displaystyle}c>{\displaystyle}l}
\tau \frac{d \delta A_1}{dt} &= & \delta   A_1 (-\mu' + i
\Delta_1) + \\ &&~~ g \left( A_0^{(s)} \delta A_2^{\ast} +
A_2^{(s)\ast} \delta A_0
 \right) +  i 2 \rho e^{i (\theta-\psi)} \delta A_2 + \\
&&~~\sqrt{2 \kappa} \delta A_1^{in} + \sqrt{2 \mu} \delta B_1^{in}\\
\tau \frac{d \delta A_2}{dt} &= & \delta   A_2 (-\mu' + i
\Delta_2) + \\ && ~~g \left( A_0^{(s)} \delta A_1^{\ast} +
A_1^{(s)\ast}\delta A_0
 \right) +  i 2 \rho e^{-i (\theta-\psi)} \delta A_1 + \\
&&~~\sqrt{2 \kappa} \delta A_2^{in} + \sqrt{2  \mu} \delta B_2^{in}\\
 \delta A_0^{(s)}& =&  \delta A_0^{in} - \frac{g}{2} \left(  A_2^{(s)} \delta A_1
+ A_1^{(s)} \delta  A_2 \right)
\end{array}\label{eq:fluct}
\end{equation}
where $\delta A_i^{in}$ and $\delta B_i^{in}$ correspond to the
vacuum fluctuations entering the cavity due respectively to the
coupling mirror and to the losses. From these equations, one can
calculate the Fourier components of the fluctuations and thus the
spectra of variances and correlations for different quadratures of
the signal and idler fields.

\section{Variances} \label{sec:variances}

The amplitude quadrature and phase quadrature fluctuations can be
written in the Fourier domain as
\begin{eqnarray}
\delta  \tilde x_i (\Omega) &=& \delta \tilde A_i(\Omega) e^{-i
\varphi_i} + \delta \tilde A_i^\ast(-\Omega) e^{i
    \varphi_i},\\
\delta \tilde p_i(\Omega) &=& -i\left(\delta \tilde A_i(\Omega)
e^{-i \varphi_i} - \delta \tilde A_i^\ast(-\Omega) e^{i
    \varphi_i}\right),
\end{eqnarray} $i=1,\, 2$ where $\delta \tilde A_i$ denotes the Fourier
component and $\varphi_i$ is the phase of the stationary solution
$A_i^{(s)}$.

To characterize the entanglement, one can calculate the normalized
variances of the sum and difference of the amplitude and phase
quadratures \cite{eprkimble,reid}
\begin{equation}
\begin{array}{>{\displaystyle}c>{\displaystyle}c>{\displaystyle}c}
S^\pm_{x}(\Omega ) &=& \frac{1}{2} \left\langle \left|\delta
\tilde x_1(\Omega ) \pm \delta \tilde x _2(\Omega )\right|^2
\right\rangle
\\S^\pm_{p}(\Omega ) &=&  \frac{1}{2} \left\langle \left| \delta \tilde p_1(\Omega )
\pm \delta \tilde p _2(\Omega )\right|^2 \right\rangle
\end{array}\label{eq:SxSp}
\end{equation}
The fact that $S_x^-$ and $S_p^+$ are smaller than one is the
signature of simultaneous quantum correlations on both phase and
amplitude quadratures. This is the case in a standard OPO above
threshold with the restriction already mentioned in the
introduction that because of the phase diffusion, $S_p^+$ can not
be measured easily.

>From the above calculated fluctuations, one can obtain the
variances for any operating point. For the sake of simplicity, we
consider here the case of detunings $\Delta = \pm 2 \rho$ which
correspond to the minimum threshold for which $\theta-\psi=
\frac{\pi}{2}$ and $A_i^{(s)} = \sqrt{\frac{2 \mu'}{g}( \sigma-1)}
e^{i \varphi_i},\, i=1,2$ where $\sigma$ is the input pump
intensity normalized to the threshold of the standard OPO. We now
have~:
\begin{eqnarray}
S^-_{x}(\Omega ) &=&  \frac{\left(\Omega ^{2}-\frac{4 \rho^{2}}{\mu
      ^{\prime 2}} \right)^{2}+\frac{4 \rho^{2}}{\mu ^{\prime
      2}}\frac{\kappa }{\mu ^{\prime }}
+\frac{\mu }{\mu ^{\prime }}\Omega ^{2}}{\left(\Omega
^{2}-\frac{4 \rho^2}{\mu ^{\prime 2}}\right)^{2}+\Omega ^{2}} \\
  S^+_{p}(\Omega ) &=&
1-\frac{\frac{\kappa }{\mu ^{\prime }}}{\Omega ^{2}+\sigma ^{2}}
\end{eqnarray}
where $\Omega =\frac{\omega \tau }{2\mu ^{\prime }}=\frac{\omega
  }{\Omega _{c}}$ is the frequency normalized to the cavity cut-off
frequency, $\Omega _{c}$.

The expression for the phase sum variance is identical to the one
obtained in the case of a standard (non phase locked) OPO~: one
obtains very large phase anticorrelations for any waveplate angle.
The expression for the intensity difference variance is different
from the expression obtained in the case of a standard OPO and
does not depend on the pump power $\sigma$. The intensity
difference noise reduction is now degraded at low frequencies by
the phase locking : inhibition of phase diffusion prevents $S^-_p$
from going to $+\infty$ as $\Omega$ goes to zero. Thus, in order
to verify the corresponding Heisenberg inequality, $S^+_p$ cannot
go to zero but instead takes a finite value.

\begin{figure}[h]
\centerline{\includegraphics[width=0.75\columnwidth,clip=]{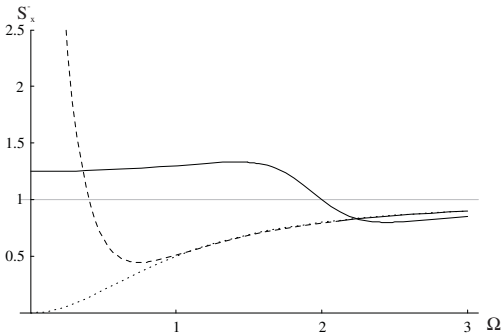}}
\caption{Intensity difference variance spectra as a function of
the normalized frequency  $\Omega$ for waveplate angles $\rho=0$
(dotted line), $\rho=0.01$, (dashed line) and
  $\rho=0.05$ (continuous line). Intracavity losses and output coupling coefficient,
  $(\mu, \,\kappa)=(0,0.05)$. The standard quantum limit is shown in grey.
These plots are independent of $\sigma$.
\label{fig:variance_intens}}
\end{figure}

\begin{figure}[h]
\centerline{\includegraphics[width=0.75\columnwidth,clip=]{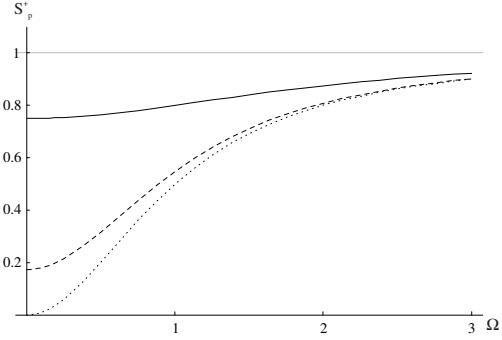}}
\caption{Phase sum variance spectra as a function of the
normalized frequency $\Omega$ for normalized pump intensity
$\sigma=1$ (dotted line), $\sigma=1.1$ (dashed line),
  $\sigma=2$ (continuous line). Intracavity losses and output coupling coefficient, $(\mu, \,\kappa)=(0,0.05)$. The
standard quantum limit is shown in grey. These plots do not depend
on the waveplate angle $\rho$.\label{fig:variance_phase}}
\end{figure}

The corresponding spectra are shown on fig
\ref{fig:variance_intens} and \ref{fig:variance_phase}.
These figures show that the amplitude quadrature difference
variance goes to zero as $\rho$ goes to zero while the phase
quadrature sum goes to zero close to threshold.

\begin{figure}[b]
\centerline{\includegraphics[width=0.75\columnwidth]{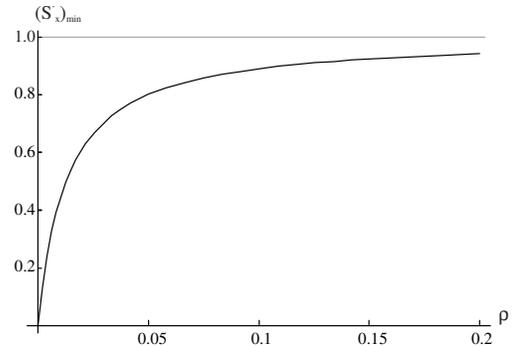}}
\caption{$\left(S^-_x\right)_{min}$ as a function of waveplate
angle $\rho$ for intracavity losses and output coupling
coefficient, $(\mu, \,\kappa)=(0,0.05)$. The standard quantum
limit is shown in grey. \label{fig:sxmoinsmin}}
\end{figure}

An important characteristic of the system is the minimal value
obtained for $S^-_x$ as a function of $\Omega$,
$\left(S^-_x\right)_{min}$ which is plotted in fig.
\ref{fig:sxmoinsmin} as a function of $\rho$.
This plot shows that one obtains a significant noise reduction as
long as $\rho$ is small compared to $\kappa$.

\section{Entanglement criteria}
\label{sec:entanglement}

\subsection{EPR correlations criterion} \label{sec:epr}

In the \textit{Gedanken Experiment} discussed in \cite{epr}, one
tries to infer the quantum state of a particle by measurements on
a spatially separated particle : the optical equivalent of that
scheme is to try to infer the signal beam by measurements on the
spatially separated idler beam. In this case, quantum correlations
do not suffice to characterize the efficiency of the system and
another criterion has to be considered \cite{reid}. The noise on
the measured observable has to be taken into account : a large
noise will reduce the precision with which the unmeasured system
is known. One is lead to introduce inference errors or conditional
variances~\cite{reid}. The conditional variances for amplitude and
phase are given by~:
\begin{eqnarray}
V_x&=& \min_g \langle|\delta \tilde x_1-g \delta \tilde x_2|^2
\rangle = \langle
|\delta \tilde x_1 - C_x \delta \tilde x_2|^2\rangle\\
V_p&=& \min_g \langle|\delta \tilde p_1-g \delta \tilde p_2|^2
\rangle = \langle |\delta \tilde p_1 - C_p \delta \tilde
p_2|^2\rangle
\end{eqnarray}
where
\begin{eqnarray}
C_x &=& \frac{\langle \delta \tilde x_1(\Omega) \delta  \tilde
x_2(-\Omega) \rangle}{\sqrt{\langle
\delta \tilde x_1^2 \rangle \langle \delta \tilde x_2 ^2\rangle }}\\
C_p &=& \frac{\langle \delta \tilde p_1 (\Omega)\delta \tilde p_2
(-\Omega) \rangle}{\sqrt{\langle \delta \tilde p_1 ^2 \rangle
\langle \delta \tilde p_2 ^2\rangle }}
\end{eqnarray}
are the normalized correlations. Ref.~\cite{reid} shows that one
must have $V_x V_p <1$ in order to have a demonstration of the EPR
paradox.

Even with correlations factors close but not equal to unity, the
conditional variances can be large if the quantum noise on the
individual beams is large as is the case for the amplitude
quadrature.

\begin{figure}[h]
\centerline{\includegraphics[width=0.75\columnwidth,clip=]{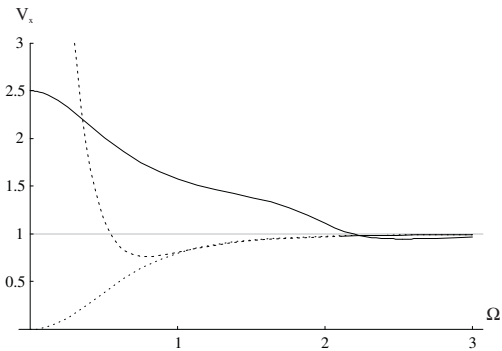}}
\caption{Conditional variance spectra for the amplitude quadrature
$V_x$ as a function of the normalized frequency $\Omega$ for wave
plate angle $\rho=0$ (dotted line), $\rho=0.01$, (dashed line) and
  $\rho=0.05$ (continuous line). Intracavity losses and output coupling coefficient,
  $(\mu, \,\kappa)=(0,0.05)$. The
  standard quantum limit is shown in grey. Normalized pump intensity
  $\sigma=1$. \label{fig:vxcond}}
\end{figure}

\begin{figure}[h]
\centerline{\includegraphics[width=0.75\columnwidth,clip=]{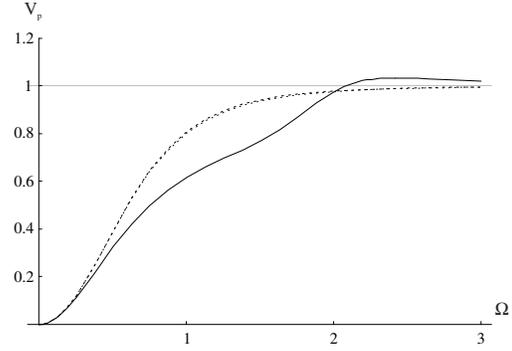}}
\caption{Conditional variance spectra for the phase quadrature
$V_p$ as a function of the normalized frequency $\Omega$ for wave
plate angle $\rho=0$ (dotted line), $\rho=0.01$, (dashed line) and
  $\rho=0.05$ (continuous line). The first two curves cannot be
  distinguished on the figure.  Intracavity losses and output coupling coefficient,
   $(\mu,\,\kappa)=(0,0.05)$. The standard quantum limit is shown in grey.
Normalized pump intensity $\sigma=1$. \label{fig:vpcond}}
\end{figure}

The conditional variances $V_x$ and $V_p$ are shown on
fig.~\ref{fig:vxcond} and \ref{fig:vpcond}. Due to the large
excess noise on the amplitude quadrature at low frequencies, $V_x$
is large for low frequencies while $V_p$ remains low for all
frequencies. This difference between the two conditional variances
is one of the main differences between our case and the other
devices which have been introduced to generate bright entangled
beams \cite{eprbachor,eprkimble,eprtaiyuan,eprpolzik}: in these
cases, one obtains theoretically symmetric behaviors for amplitude
and phase which is not the case in our system.


\begin{figure}[h]
\centerline{\includegraphics[width=.75\columnwidth,clip=]{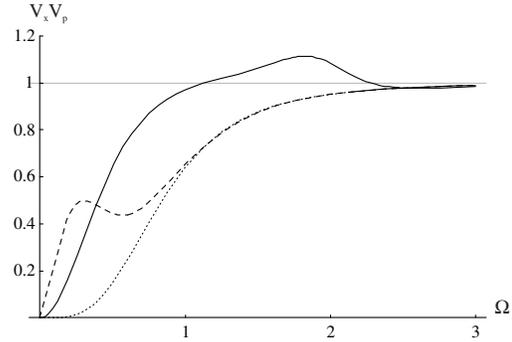}}
    \caption{$V_x V_p$  as a function of $\Omega$ for
intracavity losses and output coupling coefficient,
$(\mu,\,\kappa)=(0,\, 0.05)$ and for wave plate angle $\rho=0$
(dotted line), $\rho=0.01$, (dashed line) and  $\rho=0.05$
(continuous line). The EPR correlations limit is shown in grey.
Normalized pump intensity $\sigma=1$.\label{fig:vxvpcond}}
\end{figure}

The variation of $V_x V_p$ with $\Omega$ is plotted on
fig.~\ref{fig:vxvpcond}. This figure shows that EPR correlations
exist in a broad frequency range even for large values of the
waveplate angle. The behavior of the criterion is dominated by
that of $V_p$ which goes to zero as $\Omega$ goes to zero and is
independent of the waveplate angle.

\subsection{Inseparability criterion} \label{sec:inseparability}

However, this criterion is not always relevant. In the vast
majority of quantum information protocols, what matters is the
non-separability of the quantum state that is the impossibility to
express it as a product of independent sub-states. In order to
evaluate the inseparability of two systems, Peres proposed a
necessary criterion which applies to finite dimension systems
\cite{peres} which was latter shown by Horodecki \cite{horodecki}
to be sufficient for dimensions smaller than $2\times 2$ - or
$2\times 3$ -. In the particular case of Gaussian states, Duan
\emph{et al.} \cite{duan} and Simon \cite{simon} developed a
necessary criterion for inseparability which can be written as~:
\begin{equation}
  \label{eq:separability}
  S_x^- + S_p^+ \leq 2
\end{equation}
where $S_x^-$ and $S_p^+$ are the previously introduced variances
spectra. Let us note the entanglement produced by the self-phase
locked OPO is not in the so-called standard form (\cite{duan},
equations 10-11) : in this case, the above criterion is a
sufficient but not necessary condition for inseparability.

\begin{figure}[h]
\centerline{\includegraphics[width=.75\columnwidth,clip=]{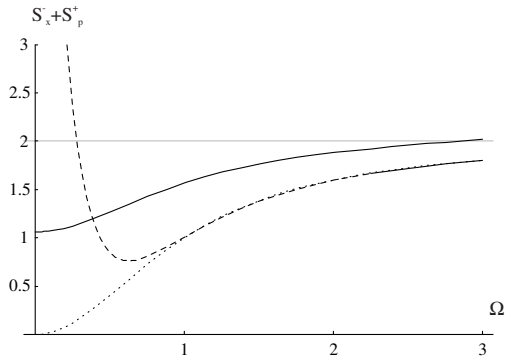}}
    \caption{$S_x^- + S_p^+$  as a function of $\Omega$ for
intracavity losses and output coupling coefficient
$(\mu,\,\kappa)=(0,\, 0.05)$ and wave plate angle $\rho=0$ (dotted
line), $\rho=0.01$ (dashed line), $\rho=0.1$ (continuous line).
The inseparability limit is shown in
    grey.\label{fig:sxplussp}}
\end{figure}

The value of $S_x^- + S_p^+$ as a function of $\Omega$ is plotted
in fig~\ref{fig:sxplussp}. At low frequencies, the behavior is
dominated by the behavior of $S_x^-$ which is large. As higher
frequencies are considered, the figure shows that the system is
indeed entangled over a large range of frequencies. When $\rho \ll
\kappa$, the excess noise on the amplitude quadrature makes the
sum larger than two meaning that the system is separable. The
curves are plotted for a value $\sigma=1$, that is at exact
threshold. Increasing $\sigma$ does not increase significantly the
variance $S_x^-$ and does not modify $S_p^+$ and will thus move
all the curves toward the inseparability limit while keeping the
sum below two. It is the behavior of the amplitude fluctuations
which dominates that of the sum of the variances.

\subsection{Fidelity of teleportation}
\label{sec:fidelity}

In teleportation experiments, one tries to transfer quantum states
from one point to another with a unity gain. In this case, the
successful of the teleportation is characterized by the overlap
between the input and output states. This overlap is usually
called the fidelity \cite{teleportkimble}. It can be shown that in
order to obtain a fidelity larger than $1 \over 2$, quantum
correlations must be used, while a fidelity larger than $2 \over
3$ denotes the possession by the receiver of the best copy
available \cite{grangier}. In the particular case of gaussian
states\footnote{and only in this case}, one can show that the
fidelity is related to the intensity difference and phase sum
variances defined in equation \ref{eq:SxSp} by the relation~:
\begin{equation}
\mathcal F = \frac{1}{\sqrt{(1+S^-_x)(1+S^+_p)}}
\end{equation}

\begin{figure}[h]
\centerline{\includegraphics[width=.75\columnwidth]{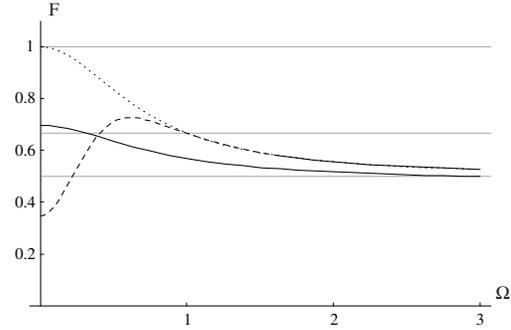}}
\caption{Fidelity, $\mathcal F$ as a function of analysis
frequency $\Omega$ for intracavity losses and output coupling
coefficient $(\mu,\,\kappa)=(0,\, 0.05)$ and wave plate angle
$\rho=0$ (dotted line), $\rho=0.01$ (dashed line), $\rho=0.1$
(continuous line). The grey lines correspond to perfect
teleportation ($\mathcal F =1$), "best copy" limit ($\mathcal F
=2/3$) and classical limit ($\mathcal F =1/2$).
 \label{fig:fidelity}}
\end{figure}

The fidelity as a function of the normalized frequency is shown in
fig \ref{fig:fidelity} for different angles of the waveplate.
One sees that for small values of the waveplate angle $\rho$, one
obtains a fidelity larger that the "best copy" limit ($\mathcal F
\geq 2/3$) over a large frequency range.

In order to determine the optimal operating parameters, one can
plot the values of waveplate angle $\rho$, normalized intensity
$\sigma$ and analysis frequency $\Omega$ for which the fidelity is
equal to the "best-copy" limit, that is $\mathcal F= \frac{2}{3}$.
The plot consists of two zones : close to threshold, \emph{i.e.}
$\sigma <1.1$, one obtains a fidelity larger than $2 \over 3$ for
large values of $\rho$ and for $\Omega < 0.5$ (dark grey surface)
; for $\sigma>1.1$, a large fidelity is observed over a larger
range of frequencies but only for very small values of $\rho$,
typically $\rho < 0.02$ (light grey surface).

\begin{figure}[h]
\centerline{\includegraphics[width=.75\columnwidth]{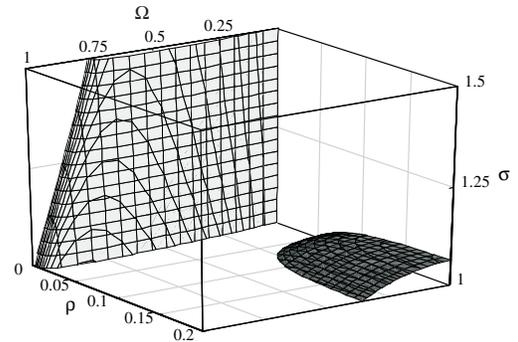}}
\caption{Values of the analysis frequency $\Omega$, wave plate
angle $\rho$ and normalized pump intensity $\sigma$ for which
$\mathcal F =\frac{2}{3}$. Intracavity losses and output coupling
coefficient $(\mu,\,\kappa)=(0,\, 0.1)$.
 \label{fig:fidelity3d}}
\end{figure}

One observes that in order to obtain a fidelity significantly
larger than $2 \over 3$, one can use very small angles, typically
less than $1^\circ$ (0.015~radians). As mentioned in ref.
\cite{part1}, this is feasible experimentally, since a small angle
does not reduce the range of the crystal temperature and cavity
length where the locking is observed but only its total surface :
once frequency degeneracy has been obtained, existing active
stabilization of the system allows for operation in this zone.
However, working very close to threshold  will reduce the
parameter range where phase locking is observed and may be
difficult experimentally.

We have shown in this section that our system produces bright
entangled beams which fulfill different entanglement criteria over
a large range of frequencies. At low frequencies, the entanglement
is observed for $\rho \ll \kappa$, that is for a rotation of the
waveplate small compared to the mirror transmission.

\section{Use of optimized quadratures}
\label{sec:opt}

The limitation of the EPR correlations zone to small angle is due
to the fact that the intensity correlations are no longer perfect
in a phase locked OPO as was the case in the standard OPO. A
simple picture shows the origin of this phenomenon. $A_1$ and
$A_2$ having a fixed phase relation, the output beam is in a fixed
polarization state which we denote $A_{bright}$. The orthogonal
polarization mode has a zero mean value and is denoted $A_{dark}$.
One can show that they correspond to the polarization eigenmodes
of the "cold" OPO, that is without the pump beam
\cite{eigenmodes}.

One can also show that their variance spectra are
\begin{equation}
\begin{array}{>{\displaystyle}c>{\displaystyle}c>{\displaystyle}c}
S_x(A_{dark}) &=& \left\langle |\delta \tilde x_{dark}|^2 \right
\rangle =
 \frac{1}{2} \left\langle |\delta \tilde x_1 - \delta \tilde x_2|^2
 \right \rangle = S_x^-\\
S_p(A_{bright}) &=& \left\langle |\delta \tilde p_{bright}|^2
\right \rangle =  \frac{1}{2} \left\langle |\delta \tilde p_1 + \delta
  \tilde p_2|^2 \right \rangle =
S_p^+
\end{array}
\end{equation}
The exact expressions of the mean fields $A_{bright}$ and
$A_{dark}$ depend on the choice of the operating point. For the
minimum threshold point where we have performed our calculations,
we have
\begin{eqnarray}
A_{bright} &=& \frac{A_1 + A_2}{\sqrt 2}\\
A_{dark} &=& \frac{A_1 - A_2}{\sqrt 2}
\end{eqnarray}
which are left and right circularly polarized due to the $\pi
\over 2$ phase shift between $A_1$ and $A_2$.

\begin{figure}[h]
\centerline{\includegraphics[width=0.75\columnwidth,clip=]{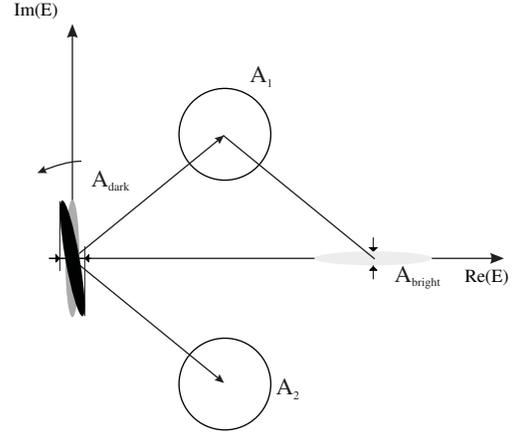}}
\caption{Fresnel picture of the signal and idler (white surfaces),
$A_{bright}$ (light grey) and $A_{dark}$ fields in the case of a
standard OPO (black) and in the case of a phase locked OPO (dark
grey). \label{fig:fresnel}}
\end{figure}

As we have seen previously that $S_x^-$ and $S_p^+$ can be smaller
than the corresponding standard quantum limit, $A_{bright}$ is
phase squeezed while $A_{dark}$ is squeezed on the orthogonal
quadrature.  This is shown in fig.~\ref{fig:fresnel} which
represents the Fresnel diagram common to the fields $A_1$, $A_2$,
$A_{bright}$ and $A_{dark}$, the phase reference being the phase
of the field $A_{bright}$.

The increase in $S_x^-$ at low frequencies corresponds to to the
tilt of the corresponding squeezing ellipse (in black) with
respect to the standard OPO situation (vertical ellipse in dark
grey). Due to the presence of the waveplate, the amplitude
quadrature of $A_{dark}$ (measurement of the width along the
horizontal axis) is not the optimal quadrature : it is thus
natural to detect this optimal quadrature instead of the amplitude
quadrature\cite{josse}. It is possible to dephase\footnote{for
instance, by using a combination of $\lambda/2$ and $\lambda/4$
waveplates.} appropriately $A_1$ with respect to $A_2$ so that
$A_{bright}$ and $A_{dark}$ are now squeezed on orthogonal
quadratures. In the simple case of the minimum threshold point,
the new entangled are simply obtained by a linear superposition of
the original fields along the neutral axis of the crystal. In the
simple case of the minimum threshold point, these fields are given
by~:
\begin{eqnarray}
A'_{1} &=& \frac{1}{\sqrt 2} \left[\cos \left( \frac{\theta_{opt}}{2}
  \right) A_1 - i \sin \left( \frac{\theta_{opt}}{2}
  \right) A_2 \right]\\
A'_{2} &=& \frac{1}{\sqrt 2} \left[\sin \left( \frac{\theta_{opt}}{2}
  \right) A_1 + i \cos \left( \frac{\theta_{opt}}{2}
  \right) A_2 \right]
\end{eqnarray}
where $\theta_{opt}$ is given by~:
\begin{equation}
\theta_{opt} = \frac{1}{2} \mathrm{arctan} \left( \frac{2 C_-
\sqrt{S_x\left(A_{dark}\right)
S_p\left(A_{dark}\right)}}{S_x\left(A_{dark}\right) -
S_p\left(A_{dark}\right)} \right)
\end{equation}
and
\begin{equation}
C_- = \frac{\left\langle\delta \tilde x\left(A_{dark}\right)
\delta \tilde p\left(A_{dark}\right)
\right\rangle}{\sqrt{S_x\left(A_{dark}\right)
S_p\left(A_{dark}\right)}}
\end{equation}\\
 The inseparability criterion developed by Duan \emph{et al.}
\cite{duan} is
\begin{equation}
  \label{eq:opt}
  S_x^+\left(A'_{bright}+\right) +
  S_p\left(A'_{dark}\right) \leq 2
\end{equation}
where $S_{x,p}\left(A_{bright,dark}\right)$ are defined above and
$A'_{bright, dark}$ are the optimally squeezed states~:
\begin{equation}
  A'_{bright,dark} = \frac{A'_1 \pm A'_2}{\sqrt 2}
\end{equation}

\begin{figure}[h]
  \begin{center}
    \includegraphics[width=.75\columnwidth]{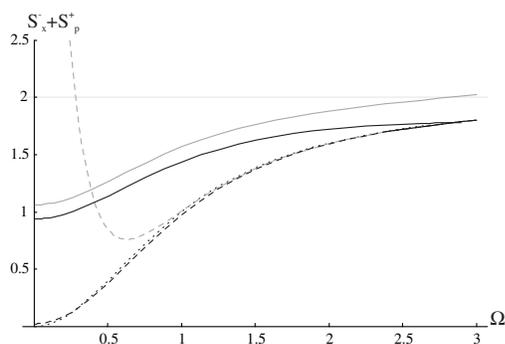}
    \caption{$S_p\left(A'_{bright}\right) +
S_x\left(A'_{dark}\right)$ as a function of analysis frequency
$\Omega$ for intracavity losses and output coupling coefficient
$(\mu,\,\kappa)=(0,\, 0.05)$ and waveplate angle $\rho=0$ (dotted
line), $\rho=0.01$ (dashed line), $\rho=0.05$ (continuous line).
Grey lines correspond to the curves obtained on fig.
\ref{fig:sxplussp} for $S_p^{+}+S_x^{-}$} \label{fig:angleopt}
  \end{center}
\end{figure}

Setting $\rho$ and $\sigma$, one can plot $S_p\left(A'_{bright}\right)
+ S_x\left(A'_{dark}\right)$ as a function of the frequency (figure
\ref{fig:angleopt}). This figure shows that one can obtain a much
larger violation of the inseparability criterion using optimized
quadratures.  For $\rho=0.01~\mbox{rad}$, the value obtained is very
close to the curve obtained for a standard OPO (where frequency non
degeneracy and phase diffusion prevent the measurement). Furthermore,
the divergence obtained at low frequencies is completely eliminated.
For larger values of the waveplate angle, the improvement is still
noticeable.

\section{Conclusion}

We have shown that a type II OPO containing a birefringent element
can be a very efficient source of intense entangled beams. This
entanglement has been characterized by the use of several
criteria, namely intensity difference and phase sum variances,
inference errors which denote the possibility to make an EPR-type
\emph{Gedanken Experiment}, inseparability criteria which
correspond to the quality of the entanglement and fidelity which
represent the quality of teleportation with unity gain.  These
criteria have been shown to be fulfilled for a large range of
operating parameters thus showing the quality of the system. The
inseparability criterion can be largely improved by the use of
optimized quadratures. We are currently setting up an experiment
based on this system.

\begin{acknowledgement} Laboratoire Kastler-Brossel, of the \'Ecole
Normale Sup\'{e}rieure and the Universit\'{e} Pierre et Marie
Curie, is associated with
the Centre National de la Recherche Scientifique.\\
This work was supported by European Community Project QUICOV
IST-1999-13071.\\
T. Coudreau is also at the P\^ole Mat{\'e}riaux et Ph{\'e}nom{\`e}nes
Quantiques FR CNRS 2437, Universit{\'e} Denis
Diderot, 2, Place Jussieu, 75251 Paris cedex 05, France.\\
We thank V. Josse and N. Treps for fruitful discussions.
\end{acknowledgement}

\end{document}